\documentclass{jetpl}
\twocolumn
\usepackage{amsmath,amsfonts,graphics,epsfig,amssymb}
\lat
\title{
Constraining the fraction of primary gamma rays at ultra-high energies
from the muon data of the Yakutsk extensive-air-shower array
}

\rtitle{
The fraction of primary gamma rays at ultra-high energies
}

\sodtitle{
Constraining the fraction of primary gamma rays at ultra-high energies
from the muon data of the Yakutsk extensive air shower array
}

\author{A.V.~Glushkov$^a$, D.S.~Gorbunov$^b$, I.T.~Makarov$^a$,
M.I.~Pravdin$^a$, G.I.~Rubtsov$^{b,c}$, I.E. Sleptsov$^a$ and
S.V.~Troitsky$^b$ }

\rauthor{A.V.~Glushkov et al.}

\address{$^a$
The Yakutsk EAS Array Collaboration,
Yu.G.~Shafer Institute of Cosmophysical Research and
  Aeronomy,
  Yakutsk 677980, Russia}

\address{$^b$ Institute for Nuclear Research of the Russian Academy of
  Sciences, 60th October Anniversary prospect 7a, Moscow 117312,
  Russia}

\address{$^c$ Faculty of Physics, M.V.~Lomonosov Moscow State
  University,
  Moscow 119992, Russia}

\dates{December 26, 2006}{*}

\abstract{
By making use of the data on the total signal and on the muon component of
the air showers detected by the Yakutsk array, we analyze, in the
frameworks of the recently suggested event-by-event approach, how large
the fraction
of primary gamma-rays at ultra-high energies can be. We derive upper
limits on the photon fraction in the integral flux of primary cosmic rays.
At the 95\% confidence level (CL), these limits are 22\% for
primary energies $E_0>4\cdot 10^{19}$~eV and 12\% for $E_0>2\cdot
10^{19}$~eV. Despite the presence of muonless events, the data are
consistent with complete absence of photons at least at 95\% CL. The
sensitivity of the results to systematic uncertainties, in particular to
those of the energy determination for non-photon primaries, is discussed.
}


\makeatletter
\renewcommand\thesection{\@arabic\c@section}

\begin{document}

\maketitle


\section{Introduction}
\label{sec:introduction}

With the increase of statistics of ultra-high-energy (UHE) cosmic-ray
(CR) events the study of the chemical composition at the very end of the
spectrum (beyond $10^{19}$~eV) is becoming quite realistic.
This issue is of primary interest today, in view of a systematic
discrepancy between energy spectra measured by different
detectors~\cite{YakutskExperiment,AGASA:energies,HiRes,Auger} and of
indications towards a fraction of neutral particles among the UHECR
primaries~\cite{neutral}. The chemical composition is also a starting
point in studies of: {\it (i) } extragalactic magnetic fields and
radiation backgrounds; {\it (ii) } accelaration mechanisms operating in
astrophysical sources;  {\it (iii) }  possible top-down scenarios emerging
in various extensions of the Standard Model of particle physics. In
particular, the photon fraction in the CR flux is of crucial importance;
the aim of this work is to derive stringent limits on this fraction in the
integral CR flux above the energy $2\cdot10^{19}$~eV.

We make use of a recently suggested
approach~\cite{Rubtsov:2006tt,Gorbunov:2006by} and perform
case-by-case analysis of 50 events detected by the Yakutsk
extensive-air-shower array (Yakutsk array in what
follows)~\cite{YakutskExperiment} with \textit{reconstructed} energies
above $2\cdot 10^{19}$~eV chosen according to quality cuts described in
Sec.~\ref{sec:data}.  To place the limit on the photon fraction, we compare
the reported information on signals measured by scintillation and muon
detectors with that expected from air-shower simulations. We focus on the
surface detector signal density at 600 meters $S(600)$ and the muon
density at 1000 meters, $\rho_{\mu}(1000)$, which are used in experiments
as primary energy and primary composition estimators, respectively.  Among
the fifty showers in the sample, two are muonless (that is, muon detectors
were operating in the shower impact area but did nor detect any signal).
These events are compatible with being initiated by primary gamma rays of
energies $2\cdot 10^{19}$~eV$<E_0<4\cdot 10^{19}$~eV, even though the
reconstructed energy exceeds $4\cdot10^{19}$~eV for one of them. One
muon-poor shower is consistent with a photon primary of energy above
$4\cdot10^{19}$~eV with probability about 10\%. For the rest of the
showers, the hypothesis of a photon primary is rejected at the 95\% CL
for each event. We derive upper limits on the fraction $\epsilon_\gamma$ of
photons in the integral flux of primary cosmic rays with \textit{actual}
energies $E_0>2\cdot 10^{19}$~eV and $E_0>4\cdot 10^{19}$~eV (the
difference between actual ($E_0$) and reconstructed ($E_{\rm est}$)
energies is discussed in Sec.~\ref{sec:data}).

The rest of the paper is organized as follows. In Sec.~\ref{sec:data}
we discuss the experimental data set used in our study. In
Sec.~\ref{sec:simulations} we briefly review the approach we use and
present our main results.  We discuss how robust these results are
with respect to changes in assumptions and in the analysis procedure
and discuss
the uncertainties associated with possible systematics in
energy determination of observed UHECR events in
Sec.~\ref{sec:robustness}. Sec.~\ref{sec:conclusions} contains our
conclusions.


\section{Experimental data}
\label{sec:data}

Yakutsk array is observing UHECR events since 1973, with detectors in
various configurations.
Since 1979, muon detectors with areas up to 36~m$^2$ (currently, five
detectors of 20~m$^2$ each with threshold energy 1~GeV for vertical
muons) supplement ground-based scintillator stations. At present, it is the
only installation equipped with muon detectors capable of studying
ultra-high-energy cosmic rays.

The energy of a primary particle is estimated from $S(600)$ and zenith
angle with the help of the procedure described in Ref.~\cite{Yakutsk_Eest},
calibrated experimentally by making use of
the atmospheric Cherenkov light. This
reconstructed energy $E_{\rm est}$ differs from the true primary
energy $E_0$ both due to natural fluctuations and due to possible
systematic effects. These latter effects depend on the primary
particle type; in particular, the difference between photons and
hadrons is significant.  Moreover, for photons, the effects of
geomagnetic field~\cite{GMF} result in directional dependence of the
energy reconstruction. Thus, the event energy reported by the
experiment should be treated with care when we allow the primary to be
a photon.  Because of possible energy underestimation for high-energy
photon-induced showers, we use events with $E_{\rm est} \ge 2\times
10^{19}$~eV even when deriving the limit for $E_0>4\cdot10^{19}$~eV;
they contribute to the final limit with different
weights~\cite{Gorbunov:2006by}.

For our study, we selected a subset of events with
$E_{\rm est} \ge 2\times 10^{19}$~eV
satisfying the following cuts aimed at the most precise determination of
both $S(600)$ and $\rho _\mu (1000)$:

{\it (i) } shower core inside the array;

{\it (ii) }  zenith angle $\theta \le 60^\circ$;

{\it (iii) } three or more muon detectors between 400~m and 2000~m
from the shower axis, operational at the moment of the shower arrival.

Our sample consists of 50 air showers; the cuts select approximately one
third of the events used for the determination of the spectrum.


\section{Simulations and results}
\label{sec:simulations}

The approach we use was described and discussed in detail in
Refs.~\cite{Rubtsov:2006tt,Gorbunov:2006by} and has already been
applied to a similar study of the photon fraction at energies above
$10^{20}$~eV~\cite{Rubtsov:2006tt}. Here, we summarise the main
steps of this approach.

For each of the events in our sample, we
generated a library of simulated showers induced by primary
photons.  Thrown energies $E_0$ of the simulated showers were randomly
selected within a relevant energy interval in order to take into account
possible deviations of $E_{\rm est}$ from $E_0$, see below. The arrival
directions of the simulated showers were the same as those of the
corresponding real events. The simulations were performed with
CORSIKA~v6.5011~\cite{CORSIKA}, choosing
QGSJET~II-03~\cite{QGSJET} as high-energy and
FLUKA~2005.6~\cite{fluka} as low-energy hadronic interaction model.
Electromagnetic showering was implemented with
EGS4~\cite{Nelson:1985ec} incorporated into CORSIKA. Possible
interactions of the primary photons with the geomagnetic field were
simulated with the PRESHOWER option of
CORSIKA~\cite{Homola:2003ru}.
As
suggested in Ref.~\cite{Thin}, all simulations were performed with
thinning level $10^{-5}$, maximal weight $10^6$ for electrons and
photons, and $10^4$ for hadrons.

For each simulated shower, we determined $S(600)$ and
$\rho_{\mu}(1000)$ by making use of the detector response functions
from Ref.~\cite{YakutskGEANT}.  For a given arrival direction, there
is one-to-one correspondence between $S(600)$ and the estimated energy
as determined by the standard analysis procedure for the Yakutsk
experiment~\cite{Yakutsk_Eest}.  This enables us to select simulated
showers compatible with the observed ones by the signal density, which
follows the Gaussian distribution in log(energy); the standard
deviation of $E_{\rm est}$ has been determined event-by-event and is
typically 17\% \cite{PravdinICRC2005}. Namely, to each simulated
shower, we assigned a weight $w_1$ proportional to this Gaussian
probability distribution in $\log E_{\rm est}$ centered at the
observed energy.  Additional weight $w_2$ was assigned to each
simulated shower to reproduce the thrown energy spectrum $\propto
E_0^{-2}$ (see Sec.~\ref{sec:robustness:spectrum} for the discussion
of the variation of the spectral index). For each of the observed
events from our dataset, we calculated the distribution of muon densities
$\rho _\mu (1000)$ representing photon-induced showers compatible with
the observed ones by $S(600)$ and arrival directions. To this end, we
calculated $\rho_\mu (1000)$ for each simulated shower by making use
of the same muon lateral distribution function as used in the analysis
of real data~\cite{Yakutsk_mu}. To take into account possible
experimental errors in the determination of the muon density, we
replaced each simulated $\rho _\mu (1000)$ by a distribution
representing possible statistical errors (Gaussian with 25\% standard
deviation~\cite{Rubtsov:2006tt}). The distribution of the simulated
muon densities is the sum of these Gaussians weighted by $w_1w_2$.

For each event we calculate,
by making use of the obtained distributions,
two numbers: the probability that it could be initiated by a photon with
true energy in the range of interest (that is,
above $E_0=4\cdot10^{19}$~eV or above $E_0=2\cdot10^{19}$~eV) and the
probability that it could be initiated by any other primary (whose energy
is assumed to be determined correctly by the experiment; see
Sec.~\ref{sec:robustness:energy} for relaxing this assumption) with
energy above this $E_0$. For most of the events, the measured muon
densities are too high as compared to those obtained from simulations
of photon induced showers.

Given these probabilities for each event, we
construct the likelihood function
(see
Ref.~\cite{Gorbunov:2006by} for details)
to estimate, at a given confidence level,
the fraction $\epsilon_\gamma$ of primary photons
among UHECR with energies in a given range. In this way we
obtain at 95\% CL
\begin{equation}
\label{main-result}
\epsilon_\gamma<22\% ~~~~{\rm ~~for~~~} E_0>4\cdot 10^{19}~{\rm
eV},
\end{equation}
\begin{equation}
\label{main-result1}
\epsilon_\gamma<12\%~~~~{\rm ~~for~~~} E_0>2\cdot 10^{19}~{\rm
eV}.
\end{equation}
These limits include corrections for the ``lost photons'' (those with
true energies $E_0>4\cdot 10^{19}$~eV for the limit~(\ref{main-result})
and $E_0>2\cdot 10^{19}$~eV for the limit~(\ref{main-result1}) but
reconstructed energies $E_{\rm rec}<2\cdot 10^{19}$~eV, see
Ref.~\cite{Gorbunov:2006by} for more details).

\begin{figure}
\includegraphics[width=\columnwidth]{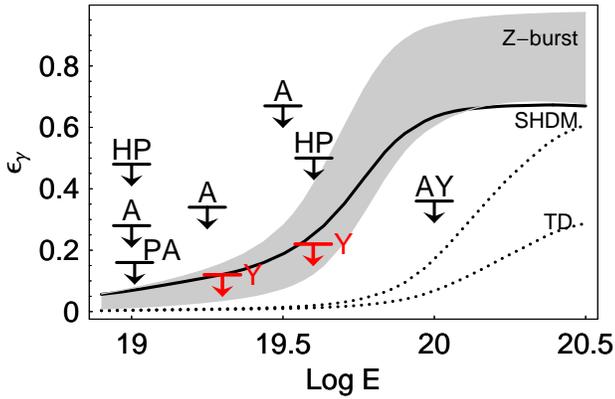}
\caption{
\label{fig:limits}
Figure 1.
Limits (95\%~CL) on the fraction
  $\epsilon_\gamma$ of photons in the integral CR flux versus
  energy. The results of the present work (Y) are shown together with
  the limits  previously given in Refs.~\cite{Haverah} (HP),
  \cite{AGASAmu} (A), \cite{Rubtsov:2006tt} (AY) and \cite{Abraham:2006ar}
  (PA). Also shown are predictions for the superheavy dark matter
  model (thick line), the topological-defect models
(necklaces, between dotted lines)~\cite{ABK}
and the $Z$-burst model (shaded area)~\cite{SS}.
  Theoretical curves are normalized to the AGASA
  spectrum~\cite{AGASA:energies}. Energy is measured in eV.}
\end{figure}

In Fig.~\ref{fig:limits},
we present our limits (denoted by Y) together with previously
published limits\footnote{A 65\% upper limit for energies above $1.2\cdot
10^{20}$~eV has been claimed from the study of AGASA
data~\cite{Risse:2005jr}; however, there are problems in accounting
for the difference between actual and reconstructed photon energies in
that work (see Ref.~\cite{Rubtsov:2006tt} for a detailed discussion).
} on $\epsilon_\gamma $.  Also, typical theoretical predictions are
shown for the superheavy dark matter, topological-defect and $Z$-burst
models. Our limits on $\epsilon_\gamma $ are currently the strongest
ones for the energy range under discussion. They disfavor the
superheavy dark matter explanation of the highest energy events.


\section{Robustness of the results}
\label{sec:robustness}
The systematic uncertainties of our results
are related to the air-shower simulations and to the data
interpretation. They were discussed in detail in
Ref.~\cite{Rubtsov:2006tt} for a different data set, with the conclusion
that the approach we use to constrain $\epsilon _\gamma $ results in quite
robust limits.

\subsection{Systematic uncertainty in the $S(600)$ and energy
determination}
\label{sec:robustness:energy}
\begin{figure}
\includegraphics[width=\columnwidth]{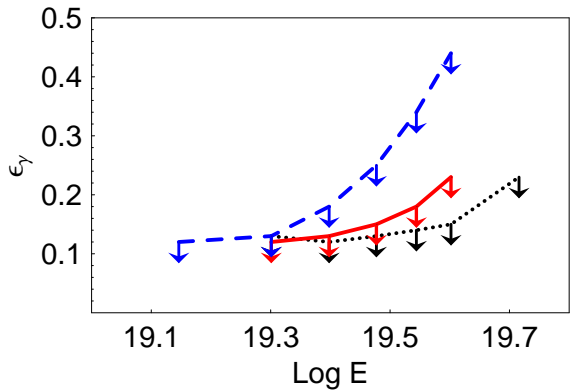}
\caption{
\label{fig:shifted}
Figure 2. Sensitivity of the limit on $\epsilon _\gamma $ to the
systematic uncertainty in the energy determination for non-photon
primaries. The solid (red) curve represents
  the limits assuming the energy scale quoted by Yakutsk experiment and
normalized to the Cherenkov light
(the same as points Y in Fig.\ref{fig:limits}). The
  dashed (blue) and dotted (black) curves correspond to the
shifts of $-30\%$ and $+30\%$, respectively, in all energies
  of non-photon primaries.}
\end{figure}

The systematic uncertainty in the absolute energy determination by the
Yakutsk array is about 30\%~\cite{YakutskExperiment}.  It originates
from two quite different sources: (a)~the measurement of $S(600)$ and
(b)~the relation between $S(600)$ and primary energy. The
probabilities that a particular event may allow for a gamma-ray
interpretation are not at all sensitive to the $S(600)$-to-energy
conversion because we select simulated events by $S(600)$ and not by
energy. These probabilities may only be affected by {\em relative}
systematics between the determinations of $\rho _\mu (1000)$ and of
$S(600)$.  On the other hand, we assumed that the experimental energy
determination is correct for non-photon primaries; the values of
probabilities that a particular event could be initiated by a
non-photon primary with energy above threshold and hence the effective
number of events contributing to the limit on $\epsilon _\gamma $
would change if the energies are systematically shifted. The effect of
such a shift would be to change the energy range for which the limit
is applicable and to change, by a few per cent, the limit itself.
This is illustrated in Fig.~\ref{fig:shifted}, which has been obtained
in a way similar to that described in Sec.~\ref{sec:simulations}, but
with six minimal values of $E_0$ for each of the three curves corresponding
to $-30\%$, $0\%$ and $+30\%$ shifts in energies of non-photon primaries.
We see that the limit at $E_0>2\cdot 10^{19}$~eV is uncertain by less
than a few per cent
while at higher energies, systematic shifts downwards reduce
statistics considerably, which results in relaxing the limit. Similar
uncertainties are expected for limits from other experiments
shown in Fig.~\ref{fig:limits}. Note that the theoretical expectations
presented there are also sensitive to the energy scale.

\subsection{Interaction models and simulation codes}
\label{sec:robustness:models}
Our simulations were performed entirely in the CORSIKA framework, and any
change in the interaction models or simulation codes, which affects either
$S(600)$ or $\rho _\mu (1000)$, may affect our limit.
As discussed in Ref.~\cite{Rubtsov:2006tt}, our method is quite
robust with respect to the changes in the interaction models and to
reasonable variations in the extrapolation of the photonuclear cross
section to high energies (the values presented here were
obtained for the standard parameterization of the photonuclear cross
section given by the Particle Data Group~\cite{PDG} and implemented as
default
in CORSIKA).

\subsection{Primary energy spectrum}
\label{sec:robustness:spectrum}
For our limit, we used the primary photon spectrum $E_0^{-\alpha }$
for $\alpha =2$.  Change in the value of $\alpha $ affects the final
limit on $\epsilon _\gamma $ through the fraction of ``lost'' photons,
but we have found that variations of $\alpha$ in the interval
$1\le \alpha \le 3$ result in variations of
$\epsilon _\gamma $ only within 1\%.

\subsection{Width of the $\rho _\mu $ distribution}
\label{sec:robustness:width}
The rare probabilities of high values of $\rho _\mu (1000)$ in
the tail of the distribution for primary photons depend on the width of
this distribution. The following sources contribute to this width:
\begin{itemize}
 \item
variations of the primary energy compatible with the observed $S(600)$
(larger energy corresponds to larger muon number and hence $\rho _\mu (1000)$);
\item
physical shower-to-shower fluctuations in muon density for a given energy
(dominated by fluctuations in the first few interactions, including
preshowering in the geomagnetic field);
 \item
artificial fluctuations in $S(600)$ and $\rho _\mu (1000)$ due to thinning;
 \item
experimental errors in $\rho _\mu (1000)$ determination.
\end{itemize}
While the first two sources are physical and are fully controlled by the
simulation code, the variations of the last two may affect the results.

It has been noted in Ref.~\cite{Badagnani} that the fluctuations in
$\rho _\mu (1000)$ due to thinning may affect strongly the precision
of the composition studies.  For the thinning parameters we use, the
relative size of these fluctuations~\cite{our-thinning} is $\sim 10\%$
for $\rho _\mu (1000)$ and $\sim 5\%$ for $S(600)$. Thus with more
precise simulations, the distributions of muon densities should become
more narrow, which would reduce the probability of the gamma-ray
interpretation of the studied events even further.

The distributions of $\rho _\mu (1000)$ we use accounted for
the error in the experimental determination of this quantity.
In principle, this error
depends on the event quality and on the muon number itself, which is
systematically lower for simulated gamma-induced showers than for the
observed events. Still, we tested the stability of our limit by taking
artificially high values of experimental errors in muon density:  50\%
instead of 25\%. The limit on $\epsilon _\gamma $ changes by less than one
per cent in that case.


\section{Conclusions}
\label{sec:conclusions}

To summarize, we have studied the possibility that ultra-high energy
events observed by the Yakutsk array were initiated by
primary photons.
The use of large-area muon detectors, a unique feature of the Yakutsk
experiment, together with the new analysis
method~\cite{Rubtsov:2006tt,Gorbunov:2006by}, enabled us to put
stringent constraints on the gamma-ray primaries even with a relatively small
set of high-quality data. An important ingredient in our study was the
careful tracking of differences between the actual and reconstructed
energies. We obtained upper bounds (\ref{main-result}),
(\ref{main-result1}) on the fraction
$\epsilon_\gamma $ of primary photons,
assuming an isotropic photon flux and $E_0^{-2}$
spectrum.
These limits are the strongest ones up to date; they
constrain considerably the
superheavy dark matter models.

We are indebted to L.G.~Dedenko and V.A.~Rubakov
for helpful discussions.
This work was supported in part by the INTAS grant 03-51-5112, by
the Russian Foundation of Basic Research grants 05-02-17363 (DG and GR),
05-02-17857 (AG, IM, MP and IS) and 04-02-17448 (DG), by the
grants of the President of the Russian Federation NS-7293.2006.2
(government contract 02.445.11.7370; DG, GR and ST), NS-7514.2006.2 (AG,
IM, MP and IS) and MK-2974.2006.2 (DG), by the fellowships of the
 "Dynasty" foundation (awarded by the Scientific Council of ICFPM, DG and
GR) and of the Russian Science Support Foundation (ST). Numerical part of
the work was performed at the computer cluster of the Theoretical Division
of INR RAS.


\end{document}